\documentclass[letterpaper,twocolumn,10pt]{article}
\usepackage{usenix-2020-09}

\usepackage{amsmath}
\usepackage{tcolorbox}

\newcommand{\system}{\textsc{MalGuard}\xspace}

\usepackage{xurl}

\usepackage{cleveref}
\crefformat{section}{§#2#1#3}

\usepackage{subcaption}

\newcommand{\PHB}[1]{\noindent\textbf{#1}\hspace{.5em}} 
\newcommand{\PHM}[1]{\vspace{.2em}\noindent\textbf{#1}\hspace{.5em}} 

\usepackage[T1]{fontenc}
\usepackage{inconsolata}
\usepackage{times}

\usepackage{bm}
\usepackage{verbatim}

\usepackage{amsthm}
\usepackage{amsfonts}
\usepackage[thinc]{esdiff}

\usepackage{pifont}

\usepackage[switch]{lineno}

\makeatletter
\newtheorem*{rep@theorem}{\rep@title}
\newcommand{\newreptheorem}[2]{%
	\newenvironment{rep#1}[1]{%
		\def\rep@title{#2 \ref{##1}}%
		\begin{rep@theorem}}%
		{\end{rep@theorem}}}
\makeatother
\newreptheorem{theorem}{Theorem}
\newreptheorem{lemma}{Lemma}
\newreptheorem{corollary}{Corollary}

\definecolor{myred}{rgb}{0.58, 0.06, 0}

\usepackage{contour}

\newcommand{\myparatight}[1]{\vspace{0mm}\noindent{\bf {#1}}~}
\usepackage[normalem]{ulem}

\contourlength{0.8pt}

\usepackage{soul}
\definecolor{mygreen}{rgb}{0.88, 0.91, 0.89}
\sethlcolor{mygreen}

\usepackage{enumitem}
\setlist{leftmargin=5mm, itemsep=0mm}

\usepackage{amsmath,graphicx}

\usepackage{numprint}
\usepackage{dsfont}  

\usepackage{balance}

\usepackage{booktabs}
\usepackage{tabularx}
\usepackage{multirow}
\usepackage{amssymb}
\usepackage{xspace}
\usepackage{xcolor}
\newcommand{\relatedrv}[1]{\textcolor{black}{#1}}

\usepackage{ulem}
\definecolor{cadmiumgreen}{rgb}{0.0, 0.42, 0.24}
\definecolor{myblue}{RGB}{0, 11, 172}
\newcommand{\sysname}{\textsc{MalGuard}\xspace}

\usepackage[absolute,overlay]{textpos}

\usepackage{soul}

\usepackage{footnote}
\usepackage{booktabs}
\usepackage[available]{usenixbadges}

\newcommand{\corresponding}{\dag}

\begin{document}

\title{\sysname: Towards Real-Time, Accurate, and Actionable Detection of \\ Malicious Packages in PyPI Ecosystem}

\author{
		\rm Xingan Gao$^{\text{1}}$ \enskip
		\rm Xiaobing Sun$^{\text{1},\corresponding}$ \enskip
		\rm Sicong Cao$^{\text{1},\corresponding}$ \enskip
		\rm Kaifeng Huang$^{\text{2}}$ \enskip
		\rm Di Wu$^{\text{3}}$ \enskip
            \rm Xiaolei Liu $^{\text{4}}$ \enskip
		\\ \rm Xingwei Lin$^{\text{5}}$ \enskip
		\rm Yang Xiang $^{\text{6}}$ 
		\\$^{\text{1}}$Yangzhou University \enskip
		$^{\text{2}}$Tongji University \enskip
		$^{\text{3}}$University of Southern Queensland 
            \\$^{\text{4}}$China Academy of Engineering Physics \enskip
            $^{\text{5}}$Zhejiang University \enskip
            $^{\text{6}}$Swinburne University of Technology
}

\pagestyle{empty}  
\maketitle

\newcommand\blfootnote[1]{%
	\begingroup
	\renewcommand\thefootnote{}\footnote{#1}%
	\addtocounter{footnote}{-1}%
	\endgroup
}

\setlength{\skip\footins}{5pt}
\blfootnote{\textsuperscript{\corresponding}Xiaobing Sun and Sicong Cao are co-corresponding authors.}


\begin{abstract}

Malicious package detection has become a critical task in ensuring the security and stability of the PyPI. 
Existing detection approaches have focused on advancing model selection, evolving from traditional machine learning (ML) models to large language models (LLMs). 
However, as the complexity of the model increases, the time consumption also increases, which raises the question of whether a lightweight model achieves effective detection. 
Through empirical research, we demonstrate that collecting a sufficiently comprehensive feature set enables even traditional ML models to achieve outstanding performance. 
However, with the continuous emergence of new malicious packages, considerable human and material resources are required for feature analysis.
Also, traditional ML model-based approaches lack of explainability to malicious packages.
Therefore, we propose a novel approach \system{} based on graph centrality analysis and the LIME (Local Interpretable Model-agnostic Explanations) algorithm to detect malicious packages.
To overcome the above two challenges, we leverage graph centrality analysis to extract sensitive APIs automatically to replace manual analysis. To understand the sensitive APIs, we further refine the feature set using LLM and integrate the LIME algorithm with ML models to provide explanations for malicious packages. 
We evaluated \system{} against six SOTA baselines with the same settings. Experimental results show that our proposed \system{}, improves \textit{precision} by 0.5\%-33.2\% and recall by 1.8\%-22.1\%. With \system{}, we successfully identified 113 previously unknown malicious packages from a pool of 64,348 newly-uploaded packages over a five-week period, and 109 out of them have been removed by the PyPI official.

\end{abstract}
\section{Introduction}
\label{sec:intro}

Python has emerged as the most popular programming language~\cite{TIOBE}.
However, threats against Python language packages have been arising in recent years~\cite{qianxin3,qianxin4}. According to a recent report~\cite{sonatype_risk} from Sonatype, a total of 704,102 malicious packages have been discovered in third-party registries by 2024, marking a year-on-year increase of more than 156\%. 
As the official package registry for Python, the Python Package Index (PyPI)~\cite{PYPI} hosts a vast array of rapidly evolving packages and their dependencies. 
Simultaneously, it has led to a growing number of security problems in the PyPI registry~\cite{DBLP:journals/corr/abs-2309-11021,Huawei, tinghua, douban, ali, tencent}. For example, a Windows Trojan ``lumma''~\cite{Lumma} targeting cryptocurrency wallets and browser extensions has extended to the PyPI registry. Attackers dropped a malicious package named \textit{crytic-compilers-0.3.9} onto the PyPI registry, intending to steal private information by typosquatting the well-known popular package \textit{crytic-compile}. 

To mitigate such security threats, an intuitive way is to regularly scan PyPI packages. Early studies primarily employed static analysis and/or Machine Learning (ML)~\cite{DBLP:conf/icse/SejfiaS22,DBLP:journals/corr/abs-2309-02637,virustotal,FORK,Microsoft2,DBLP:conf/kbse/LiangLWLW23,DBLP:conf/ndss/DuanAKESL21} to identify malicious features.
However, hand-crafted expert rules are hard to obtain and difficult to keep pace with the rapidly evolving malicious behaviors~\cite{DBLP:journals/corr/abs-2309-11021}. 
To improve the usability of the existing approaches and avoid the intense labor of human experts on feature extraction, recent works investigate the potential of Large Language Models (LLMs) in a more automated way of malicious package detection \cite{DBLP:journals/corr/abs-2309-02637,sun2024,2024Shifting}. For example, Zhang et al. \cite{DBLP:journals/corr/abs-2309-02637} fine-tuned the BERT model~\cite{DBLP:conf/naacl/DevlinCLT19} to understand the semantics of sequential malicious behaviors.
 
Despite their effectiveness, fine-tuning a dedicated LLM is accompanied by considerable computational overhead, while directly invoking commercial LLMs (e.g., ChatGPT) via API will entail significant deployment cost when dealing with massive packages within the PyPI ecosystem \cite{DBLP:journals/tosem/CaoSBWLWTZL24}. Therefore, it motivates us to explore a simple yet effective approach that can strike a balance between effectiveness and overhead, aligning with the requirements of real-time and accurate detection needed in industrial settings.

To this end, we first performed an empirical study that aims to investigate \emph{whether simple ML models can achieve comparable effectiveness to LLMs in malicious package detection}.
First, we constructed a feature set of 132 dimensions. Using this feature set, we performed feature extraction and trained five traditional ML models. i.e., Random Forest (RF), Naive Bayes (NB), Extreme Gradient Boosting (XGBoost)~\cite{DBLP:conf/kdd/ChenG16}, Multilayer Perceptron (MLP), and Support Vector Machine (SVM). Our experimental results show that the RF model performs the best among all five ML models that we selected.
While the \textit{precision} is slightly lower than \textsc{Ea4mp}, \textsc{Cerebro}, and GPT-3.5-turbo~\cite{ChatGPT}, the difference is negligible. Importantly, the RF model exhibits significantly higher \textit{recall} compared to these two state-of-the-art (SOTA) approaches.
It indicates that with a comprehensive feature set, we can achieve comparable detection accuracy with a lightweight model. 
Furthermore, while pre-training or fine-tuning LLMs typically requires several hours to even tens of hours, it takes approximately 5 seconds to train ML models and less than 1 second to perform malicious detection per package. The high efficiency enables ML model-based malicious package detection to perform real-time detection on newly updated packages. 

Nevertheless, we summarize two remaining challenges for ML models that can hinder the effectiveness of malicious package detection:

\begin{itemize}[leftmargin=1em]
    \item \textbf{Challenge 1 (C1): Reducing Dependency on Manually Pre-defined Feature Sets.}  
    Quality features lead to better performance, but existing approaches often rely on manual analysis of malicious packages to construct these feature sets. 
    As time progresses, the dataset is continuously augmented with new malicious package samples, necessitating ongoing manual effort from security professionals to analyze their characteristics.
    This process is resource-intensive and may miss critical features.
    
    \item \textbf{Challenge 2 (C2): Generate Explainable Outputs of Suspicious Behaviors for Malicious Packages.}  
    Current detection efforts tend to focus on binary classification tasks for identifying malicious packages. However, once a malicious package is identified, administrators must spend significant time manually verifying its behavior. 
    Although GuardDog~\cite{guarddog} provides descriptions of suspicious APIs found in malicious packages, its rule-based approach requires extensive manual effort to define specific rules and lacks comprehensive analysis of the APIs \cite{coca}. Further assistance is needed to help administrators understand which sensitive APIs are being used in malicious packages and to identify their potential malicious purposes~\cite{explannation}, which has received little attention in prior work.
\end{itemize}

To address the two challenges, we propose a novel malicious package detection approach \system{}. 

To address \textbf{C1}, we employ static analysis methods to construct API call graphs for each malicious package and compute the centrality value of individual APIs. We then average the centrality values of APIs with the same name across different samples. Based on these averaged scores, we rank the APIs and select the top-$K$ as the sensitive API feature set. To enhance the \textit{precision} of the identified sensitive APIs, we employ role-based prompt engineering and utilize LLM (GPT-3.5-turbo) to assess the top-$K$ candidates.
To address \textbf{C2}, we leverage LLM to analyze each sensitive API and infer its potential malicious purposes. This resulted in a ground truth dataset containing suspicious API names and their potential malicious behaviors. Notably, our approach requires only one query to the LLM per sensitive API. The feature set is incrementally updated as new sensitive APIs emerge.

Using this feature set, we extracted the feature vector from the dataset and trained ML models integrated with the LIME algorithm.
The algorithm identifies the top 10 influential non-zero features in the model's decision-making process. These features are then cross-referenced with the \textit{\{API\_name : API\_malbehavior\}} Ground\_truth dataset to generate detailed explanation outputs for each malicious package. These outputs provide a clear explanation of the malicious behavior associated with the identified sensitive APIs for researchers to verify these packages.

Prior approaches rely on manually refined feature sets. In comparison, our approach leverages graph centrality analysis and automates feature extraction with the assistance of LLMs, eliminating the need for manually predefined feature sets. In addition, we incorporate the LIME algorithm to generate explanations for the malicious packages.

Moreover, our work involves more than just simple name-based retrieval operations. We also sorted the APIs by their invocation order within each file and systematically output the content of these explanations. This ordered presentation facilitates a better understanding of malicious packages for researchers, providing them with clearer insights into the sequence and context in which APIs are used.

We evaluated \system{} against six SOTA baselines on the new dataset, which consists of malicious packages collected from Guo et al.~\cite{DBLP:journals/corr/abs-2309-11021} and Sun et al.~\cite{sun2024}. Experimental results show that \system{} improves \textit{precision} by 0.5\%-33.2\% and \textit{recall} by 1.8\%-22.1\%. We monitored 64,348 software packages uploaded on PyPI between December 21, 2024, and January 28, 2025, and successfully identified 113 previously unknown malicious packages. We reported these malicious packages to PyPI officials, and 109 of them have been removed.

The contributions of our paper are as follows:
\begin{itemize}[leftmargin=1em]
\item We propose \system{}, a novel approach based on Social-network Graph Centrality to detect malicious PyPI packages.
\item We collected a feature set containing 132 features to conduct our empirical study and demonstrate that, with a sufficiently comprehensive feature set, lightweight models can achieve effectiveness comparable to LLMs.
\item  We use LLMs and ML models to achieve explainability in malicious package detection.
\item \system{} has uncovered 113 previously unknown malicious packages. We reported these packages to PyPI officials, and 109 of them have been removed. 

\end{itemize}

\section{Empirical Study}
\label{sec:background_es}

Since attacks on the PyPI platform are ongoing and continuously evolving, it is necessary to update detection models with newly emerging malicious samples. Notably, existing SOTA approaches, such as \textsc{Cerebro} and \textsc{Ea4mp}, rely on fine-tuning large pre-trained models, and such fine-tuning incurs substantial time and computational costs. In this section, we conduct two empirical studies. First, we investigate whether lightweight ML models can achieve detection effectiveness comparable to SOTA approaches. Second, we examine how temporal differences (i.e., the time gap between training and test samples) affect models' effectiveness.

\begin{table}[t]
 \caption{Statistics of the constructed dataset.}
  \centering
  \resizebox{\columnwidth
  }{!}{
  \begin{tabular}{c|cc}
    \toprule
    \textbf{Dataset} &\textbf{\#Malicious} &\textbf{\#Benign}\\
    \midrule
    Guo et al. ~\cite{DBLP:journals/corr/abs-2309-11021} & 9,148 & - \\
    Sun et al. ~\cite{sun2024}& 516 & -  \\
    Our work & - & 10,000 \\
    \midrule
    Total & 9,664 & 10,000 \\
    \bottomrule
  \end{tabular}
  }
  \label{tab:malicious}
\end{table}

\subsection{Dataset} 

\label{sec:background_dataset}

\PHM{Malicious Sample.} Since the detection capability of learning-based approaches benefits from large-scale and high-quality datasets, we built our evaluation benchmark by merging two reliable human-labeled datasets collected from real-world Python packages, including Guo et al.~\cite{DBLP:journals/corr/abs-2309-11021} and Sun et al.~\cite{sun2024}. 
Detailed statistics for the two datasets are shown in Table \ref{tab:malicious}.
In total, our merged dataset contains 9,664 malicious PyPI packages.

\noindent \PHM{Benign Sample.} We randomly sampled 10,000 popular packages from PyPI. Following \cite{DBLP:conf/kbse/LiangLWLW23,sun2024}, a package will be considered as benign if it has been (\ding{182}) hosted on PyPI for more than 90 days and (\ding{183}) downloaded over 1,000 times.

\begin{table}[htbp]
  \centering
  \caption{The categories of 132 different APIs in Feature Set.}
  \resizebox{0.8\columnwidth
  }{!}{
    \begin{tabular}{c|c}
    \toprule
    Categories & API example \\
    \midrule
    \multirow{5}[2]{*}{File-system access} 
      & os.mkdir() \\
      & os.remove \\
      & shutil.copy() \\
      & write() \\
      & ... \\
    \midrule
    \multirow{4}[2]{*}{Process creation} 
      & subprocess.Popen \\
      & multiprocessing.Process \\
      & threading.Thread \\
      & ... \\
    \midrule
    \multirow{4}[2]{*}{Network access} & socket.socket() \\
      & requests \\
      & request.urlopen() \\
      & ... \\
    \midrule
    \multirow{3}[2]{*}{Data encode \& decode} & base64.b64encode() \\
      & base64.b64decode() \\
      & ... \\
    \midrule
    \multirow{3}[2]{*}{Package install} & install.run() \\
      & pip.main() \\
      & ... \\
    \midrule
    \multirow{3}[2]{*}{System access} & os.getenv() \\
      & os.getcwd() \\
      & ... \\
    \bottomrule
    \end{tabular}%
    }
  \label{tab:categories}%
\end{table}%

\begin{table*}[htbp]
  \centering
  \caption{Effectiveness comparison of five different ML models and LLM-based approaches on the same dataset.}
 \resizebox{2\columnwidth
 }{!}{
    \begin{tabular}{|c|c|c|c|c|c|c|}
    \toprule
    \multirow{2}[4]{*}{Group} & \multirow{2}[4]{*}{Model} & \multirow{2}[4]{*}{Precision (\%)} & \multirow{2}[4]{*}{Recall (\%)} & \multirow{2}[4]{*}{F1 score (\%)} & \multicolumn{2}{c|}{Time Consumption} \\
\cmidrule{6-7}      &   &   &   &   & Pre-process (s/package) & Train (s) \\
    \midrule
    \multirow{5}[10]{*}{ML} & NB & 55.2 & 98.4 & 70.7  & \multirow{5}[10]{*}{\textbf{0.8457}} & 0.19467 \\
\cmidrule{2-5}\cmidrule{7-7}      & XGBoost & 98.1 & 98.4 & 98.2  &   & 4.79 \\
\cmidrule{2-5}\cmidrule{7-7}      & RF & 98.5 & 98 & 98.2  &   & 1.0126 \\
\cmidrule{2-5}\cmidrule{7-7}      & SVM & 89.2 & 94.7 & 91.9  &   & 0.097 \\
\cmidrule{2-5}\cmidrule{7-7}      & MLP & 98.1 & 98.2 & 98.1  &   & 22.85157 \\
    \midrule
    \multirow{2}[4]{*}{PTM} & \textsc{Ea4mp}~\cite{sun2024} & \textbf{99.1} & 95.4 & 97.2  & 6.28 & 30,741.67 \\
\cmidrule{2-7}      & \textsc{Cerebro}~\cite{DBLP:journals/corr/abs-2309-02637} &  98.6 & 85.7  & 91.7  &  12.489 & 2,439 \\
    \midrule
    LLM & GPT-3.5-turbo~\cite{ChatGPT} & 99.0 & \textbf{99.3} & \textbf{99.1}  & - & -  \\
    \bottomrule
    \end{tabular}%
    }
  \label{tab:result1}%
\end{table*}%

\subsection{Baseline}
\label{sec:background_llms}
To evaluate the differences between LLMs and ML models in malicious package detection, we compared our approach with two of the latest SOTA approaches: Zhang et al.~\cite{DBLP:journals/corr/abs-2309-02637} proposed \textsc{Cerebro}, an approach for extracting code behavior sequences using abstract syntax trees (AST). By analyzing the AST, they extracted available APIs to form code sequences that describe malicious behaviors. These sequences were then used as input for fine-tuning a BERT model. Similarly, Sun et al.~\cite{sun2024} introduced an integrated detection approach based on deep code behavior sequences and metadata. Their approach used static analysis tools to extract control-flow graphs (CFGs) and call graphs (CGs) to generate code behavior sequences, which were subsequently fine-tuned with the BERT model. We leveraged the one-shot technique along with role-based promoting to analyze the test dataset\footnote{For cost considerations, we randomly selected 300 benign packages and 300 malicious packages from the dataset for testing when conducting experiments with ChatGPT.} using GPT-3.5-turbo~\cite{ChatGPT}. All these approaches have demonstrated exceptional effectiveness in their respective datasets, so we chose them as the benchmarks for comparison.

\subsection{Study 1: Effectiveness Comparison of ML Models and LLM-based Approaches}
\noindent \PHM{Experiment Setup.} 

To evaluate whether traditional ML models can achieve competitive effectiveness compared to LLM-based approaches in malicious package detection, we conducted experiments using five widely adopted ML classifiers: Extreme Gradient Boosting (XGBoost), Random Forest (RF), Naive Bayes (NB), Support Vector Machine (SVM), and Multi-Layer Perceptron (MLP).
\begin{table}[htbp]
  \centering
  \caption{Temporal partitioning of the dataset.}
    \begin{tabular}{c|cccc}
    \toprule
    Year & 2021 & 2022 & 2023 & 2024 \\
    \midrule
    Counts & 227 & 1,054 & 6,412 & 1,971 \\
    \bottomrule
    \end{tabular}%
  \label{tab:temporal_order}%
\end{table}%
To ensure the quality of the feature set, we manually analyzed 9,664 malicious packages and incorporated features identified in previous studies. This process yielded a comprehensive feature set with 132 dimensions. These features were further organized into six distinct categories, as summarized in Table~\ref{tab:categories}.

\noindent \PHM{Result.} 
The experimental results are shown in Table~\ref{tab:result1}. Four ML (i.e., XGBoost, RF, SVM, and MLP) models exhibited effectiveness comparable to LLM-based approaches. Among them, Random Forest (RF) and XGBoost models performed particularly well, achieving \textit{precision} and \textit{recall} rates of 98.0\% and 98.2\%, respectively. The Naive Bayes (NB) model yielded a lower \textit{precision} of 55.2\%, underperforming than LLM-based methods. GPT-3.5-turbo achieved the highest \textit{recall} and \textit{F1 score} among all approaches, highlighting the inherent advantage of large language models in understanding code semantics. Nonetheless, the effectiveness gap is not substantial. For instance, the Random Forest model achieved an accuracy only 0.5\% lower and a recall rate just 1.3\% lower than that of GPT-3.5-turbo. Compared to \textsc{Ea4mp} and \textsc{Cerebro}, while their \textit{precision} (slightly below \textsc{Ea4mp}'s 99.1\% and \textsc{Cerebro}`s 98.6\%) is marginally lower, ML models' recall and \textit{F1 scores} are significantly higher.

Moreover, in terms of time efficiency, we found that for ML models, feature vector extraction can be achieved using simple static analysis tools or even regular expression matching, requiring an average of only 0.0035 seconds per package. Training an ML model takes 0.097 to 22.85 seconds, which is significantly faster compared to \textsc{Ea4mp}'s 6.28 seconds and \textsc{Cerebro}`s 12.489 seconds per package and time required to fine-tune a BERT model. This highlights the significant time advantage of ML models.

\begin{table*}[htbp]
  \centering
  \caption{Effectiveness comparison of different ML models and LLM-based approaches on newer samples by training an old dataset.}
  \resizebox{2.0\columnwidth
  }{!}{
    \begin{tabular}{r|rrr|rrr|rrr|rrr|rrr}
    \toprule
      & \multicolumn{3}{c|}{XGBoost} & \multicolumn{3}{c|}{RF} & \multicolumn{3}{c|}{SVM} & \multicolumn{3}{c|}{MLP} & \multicolumn{3}{c}{EA4MP} \\
    \midrule
       \multicolumn{1}{l|}{Metrics (\%)}& \multicolumn{1}{l}{Precision} & \multicolumn{1}{l}{Recall} & \multicolumn{1}{l|}{F1} & \multicolumn{1}{l}{Precision} & \multicolumn{1}{l}{Recall} & \multicolumn{1}{l|}{F1} & \multicolumn{1}{l}{Precision} & \multicolumn{1}{l}{Recall} & \multicolumn{1}{l|}{F1} & \multicolumn{1}{l}{Precision} & \multicolumn{1}{l}{Recall} & \multicolumn{1}{l|}{F1} & \multicolumn{1}{l}{Precision} & \multicolumn{1}{l}{Recall} & \multicolumn{1}{l}{F1} \\
    \midrule
    \multicolumn{1}{l|}{2021\&2022} & 88.2  & 80.3  & 84.1  & 97.1  & 82.0  & 88.9  & 88.6  & 80.3  & 84.2  & 95.3  & 80.6  & 87.3  & 94.7  & 90.7  & 92.7  \\
    2023 & 86.4  & 59.0  & 70.1  & 90.1  & 59.3  & 71.5  & 83.3  & 49.2  & 61.9  & 87.3  & 62.1  & 72.6  & 81.6  & 84.3  & 82.9  \\
    2024 & 81.5  & 53.4  & 64.5  & 72.6  & 52.1  & 60.7  & 75.4  & 51.0  & 60.8  & 79.6  & 57.1  & 66.5  & 72.7  & 70.5  & 71.6  \\
    \bottomrule
    \end{tabular}%
    }
  \label{tab:scenario_1}%
\end{table*}%

\subsection{Study 2: Robustness Against PyPI Malicious Packages}
\noindent \PHM{Experiment Setup.} To evaluate the resilience of existing approaches and ML models in malicious package detection, we partitioned the malicious package dataset chronologically, as detailed in Table~\ref{tab:temporal_order}. To ensure the validity of the evaluation, we selected an equal number of popular benign packages from each year, maintaining a 1:1 ratio between benign and malicious samples. Given that the number of malicious packages in 2021 was relatively limited and significantly fewer than in subsequent years, we merged the samples from 2021 and 2022 to form the training set. The models were then evaluated separately on samples from 2023 and 2024. Based on the newly defined training set, we reanalyzed the feature distributions and reconstructed the feature set. For ML models, we selected four classifiers that demonstrated strong effectiveness in Study 1: XGBoost, RF, SVM, and MLP. Since we are unable to locally fine-tune or deploy GPT-3.5-turbo, we chose \textsc{Ea4mp}, which achieves effectiveness second only to GPT-3.5-turbo, as the baseline for comparison.

\noindent \PHM{Result.} The experimental results, as shown in Table~\ref{tab:scenario_1}, demonstrate that with the addition of new samples, models trained solely on outdated data experience a notable effectiveness decline. For instance, in the case of ML models, the XGBoost model shows only a modest drop in \textit{precision} (from 88.2\% to 81.5\%), yet its recall plummets significantly from 80.3\% to 53.4\%, indicating that a large portion of newly uploaded malicious samples cannot be effectively detected by the old model. Similarly, the \textit{F1 score} of \textsc{Ea4mp} decreases from 92.7\% to 71.6\%, reflecting a degradation in its detection capability.

Combined with the findings from \textbf{Study 1}, it is evident that both ML-based and LLM-based approaches require continuous updates with new data to maintain detection effectiveness. Compared to LLM-based approaches, ML models offer the advantage of faster iteration. Moreover, it is important to emphasize that while ML models may achieve detection effectiveness comparable to LLMs, constructing a comprehensive and high-quality feature set remains a major challenge. In our study, three master's students spent over a week analyzing 9,664 malicious packages to develop the final feature set consisting of 132 dimensions, amounting to 21 person-hours. Therefore, it is essential to develop a solution that supports efficient and automated feature set updates to enable timely and scalable model iteration.

\section{API Call Graph Based Centrality Analysis}
\label{sec:background}

To enable automated extraction and rapid iteration of the feature set, we draw inspiration from Android malware detection, where graph centrality analysis has proven to be effective in identifying frequently invoked sensitive APIs. In this section, we provide a detailed description of the API Call Graph and the application of centrality analysis. Furthermore, we investigate whether there exist significant differences in API invocation patterns between benign and malicious packages within the PyPI ecosystem.

\subsection{API Call Graph and Centrality Analysis}
\noindent \PHM{API Call Graph.} The Application Programming Interface (API) Call Graph is a commonly used data structure in static or dynamic program analysis \cite{SP}, designed to abstract the calling relationships among various API functions within a software system. In such a graph, each node represents an API function, method, or module, while edges denote the direction of invocation (i.e., the source node calls the target node). Compared to traditional structures such as the Control Flow Graph (CFG), the API Call Graph focuses more on high-level semantic invocation logic. API call graphs have been extensively employed in the security domain for malware detection~\cite{sun2024,intdroid,maltracker}, where they facilitate the identification of potential threats through the analysis of abnormal invocation patterns and frequent co-occurrence of malicious API sequences.

\noindent \textbf{Centrality Analysis.} Centrality metrics commonly used in social network analysis, such as degree centrality~\cite{freeman2002centrality}, Katz centrality~\cite{katz1953new}, closeness centrality~\cite{freeman2002centrality}, and harmonic centrality~\cite{marchiori2000harmony}, have been widely applied in Android malware detection tasks~\cite{malscan, intdroid}. Studies have shown that these metrics effectively capture the behavioral characteristics of critical nodes within malicious code, thereby enhancing the discriminative power of detection models. Structurally, API call graphs share fundamental modeling similarities with social network graphs in both form and graph-theoretic properties. Both can be formalized as directed graphs, where nodes represent individual entities (typically users or actors in social networks, and functions, methods, or class modules in API call graphs), while edges denote interactions, such as social connections in the former and invocation or execution dependencies in the latter. These graph types also exhibit similar topological properties. For example, their node degree distributions often follow a power-law pattern, where a small number of nodes (e.g., frequently invoked functions or influential users) possess disproportionately high connectivity, while most nodes remain peripheral. Moreover, both types of graphs tend to form local clusters or community structures—subgraphs where node connectivity is significantly denser than the global average. These structures often correspond to functionally cohesive modules (such as attack chains) or social communities.

\subsection{Difference  between Malicious and Benign Packages}
\label{subsec:differ}
To verify whether malicious PyPI packages share similar traits with Android malware, we constructed API call graphs for both benign and malicious packages in the dataset and calculated the centrality values of APIs to compare their differences.
The Experiment results highlight significant differences in API usage tendencies between benign and malicious packages. For instance, in the malicious package dataset, APIs such as \textit{exists}, \textit{subprocess.Popen}, \textit{os.getenv}, \textit{install.run}, \textit{b64decode}, and \textit{encode} are frequently invoked. Attackers often use \textit{os.getenv} to evade dynamic analysis tools, \textit{subprocess.Popen} to create malicious processes, and advanced attackers may employ APIs like \textit{b64decode} to obfuscate data, making detection more challenging. In contrast, benign packages tend to favor simpler APIs for data processing, such as \textit{int}, \textit{str}, \textit{list}, and \textit{print}. Due to the limitation of the length of the article, we only present the top 10 APIs ranked by centrality scores calculated for both benign and malicious package datasets using four different centrality metrics in Table~\ref{tab:MOTIVATION}.

\begin{table}[htbp]
  \centering
  \caption{The top 10 APIs calculated with different centrality in malicious\&benign packages.}
  \resizebox{\columnwidth
  }{!}{
    \begin{tabular}{c|l|l|l|l}
    \toprule
      & Closeness & Degree & Harmonic & Katz \\
    \midrule
    \multirow{10}[2]{*}{Malicious} & setup & setup & join & setup \\
      & exists & exists & open & exists \\
      & subprocess.Popen & subprocess.Popen & decode & subprocess.Popen \\
      & open & join & getattr & open \\
      & join & open & encode & join \\
      & range & range & map & install.run \\
      & getattr & aetattr & exists & exec \\
      & map & map & os.getenv & format \\
      & os.getenv & exec & replace & os.getenv \\
      & install.run & os.getenv & b64decode & expanduser \\
    \midrule
    \multirow{10}[2]{*}{Benign} & open & open & len & setup \\
      & len & len & join & open \\
      & setup & setup & str & len \\
      & print & join & isinstance & join \\
      & str & print & open & print \\
      & isinstance & str & int & str \\
      & int & isinstance & list & isinstance \\
      & format & int & print & int \\
      & list & range & append & range \\
      & super & format & super & list \\
    \bottomrule
    \end{tabular}%
    }
  \label{tab:MOTIVATION}%
\end{table}%

To address these discrepancies, we propose a detection tool based on API call graph centrality and ML models. For each malicious package, we generate an API call graph and compute the centrality value for its nodes. By aggregating and averaging the centrality scores of nodes with the same API name across all malicious packages, we rank the APIs based on their averaged centrality scores. We then select the top $K$ ($K$\( = 200, 300, 400, 500\))  APIs as the feature set for feature vector extraction. Using these extracted feature vectors, we train an ML model to detect malicious packages effectively.

\section{\system{}: Graph Centrality and ML-Based Malicious Package Detection}~\label{sec:impl}

\begin{figure*}[t]
	\centering
	\includegraphics[width=2\columnwidth]{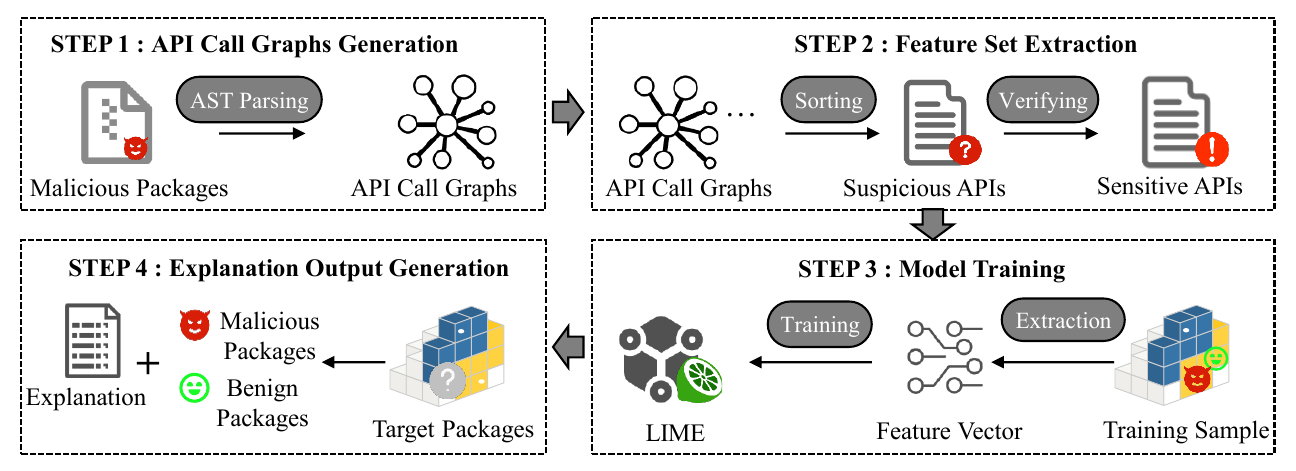}
	\caption{\system{} architecture.}
	\label{fig:overflow}
\end{figure*}

To automate the extraction of sensitive API feature sets and train ML models for malicious package detection,  and improve the explainability of ML models, we propose a novel approach: \system{}. The workflow of \system{}, illustrated in Figure~\ref{fig:overflow}, comprises Four main steps:
\textbf{Step 1: API Call Graph Generation.}  We perform static analysis on each malicious package to extract its AST. Based on the AST, we generate an API call graph and calculate the centrality values of its nodes. 
\textbf{Step 2: Sensitive API Feature Set Extraction and Filter.} We compute the centrality values of all nodes across the malicious packages, summing, averaging, and ranking the values for APIs with the same name. The top 500 APIs with the highest centrality values are selected as the sensitive API feature set. 
Based on the feature set that was extracted, we leveraged the one-shot technique along with role-based promoting to send the extracted APIs to an LLM (\textit{GPT-3.5-turbo}) for analysis. If the API could potentially be used for malicious purposes, we retained it in the feature set; otherwise, it was removed. Additionally, for the retained APIs, we leveraged the language model to perform an analysis of possible malicious behaviors. The analysis results were saved in the format \textit{{api\_name: malicious\_behavior}}, creating a Ground\_Truth dataset for further reference.
\textbf{Step 3: Model Training.} Using the feature set obtained in the second step, we extract feature vectors and train an ML model for malicious package detection. 
\textbf{Step 4: The Explanation Output based on LIME Algorithm.} Based on the ML model trained in Step 3, we integrated the LIME algorithm into the model. By identifying the top 10 most influential non-zero features for the model's decision, we matched these feature names against the Ground\_Truth dataset. For each matched feature, all its associated potential malicious behaviors were retrieved and compiled to produce an explanation output, providing insights into the reasoning behind the model's detection decision.

\subsection{API Call Graph} Generation
Since social network graphs have been proven effective in Android malware detection, and our experiments in Section~\ref{subsec:differ} further demonstrate that sensitive APIs in PyPI malicious packages often exhibit significantly higher and more anomalous centrality values in API call graphs compared to those in benign software, we focus our analysis on malicious packages when extracting API call graphs.  

First, for each malicious package, we construct an AST. Using the relationships between nodes and edges in the AST, we generate a corresponding API call graph for each package. In the API call graph, each node represents an API, and the edges capture the invocation relationships between APIs. Based on these relationships, we calculate the centrality values for all nodes (APIs).  

Given the variety of centrality measures available, we selected four of the most widely used centrality metrics to ensure comprehensive analysis: \textbf{Closeness centrality}, \textbf{Degree centrality}, \textbf{Katz centrality}, and \textbf{Harmonic centrality.} These measures provide a framework for identifying APIs with anomalous behaviors that may indicate malicious activity.

\textbf{Closeness centrality.} Closeness centrality is based on the reciprocal of the sum of the shortest path distances from a node to all other nodes, emphasizing the node's efficiency in spreading information or influence.

\[
C_{C}(v)=\frac{N-1}{\sum_{u \in V, u \neq v} d(v, u)}
\]
Where \(N\) is the total number of nodes in the graph, \(d(v,u)\) represents the shortest path distance between node \(v\) and \(u\),\(V\) is the set of all nodes in the graph.

\textbf{Degree centrality.} Degree centrality evaluates a node's importance based on the number of direct connections (edges) it has with other nodes. 

\[
C_{D}(v)=\frac{\operatorname{deg}(v)}{N-1}
\]
Where \(N\) is the total number of nodes in the graph, \(\operatorname{deg}(v)\) represents the degree of the node \(v\) (\textit{i.e.}, the number of edges connected to node \(v\)) and \(N\) is the total number of nodes in the graph.

\textbf{Katz centrality.} Katz centrality measures a node's influence by considering both the immediate neighbors and the neighbors further away, applying a weighting factor to penalize more distant connections.

\[
C_{K}(v)=\alpha \sum_{u \in V} A_{v u} C_{K}(u)+\beta
\]

Where \(\alpha\) is the damping factor, typically a small value less than 1, which controls the influence of distant nodes, \(A_{v u}\) is the entry in the adjacency matrix \(V\) (\(A_{v u}=1\) if there is an edge from node \(v\) to \(u\), otherwise (\(A_{v u}=0\)), \(\beta\) is a constant that can be adjusted to represent the baseline centrality of each node.

\textbf{Harmonic centrality.} Harmonic centrality accounts for disconnected nodes by summing the reciprocal of distances rather than taking their total sum.

\[
C_{H}(v)=\sum_{u \in V, u \neq v} \frac{1}{d(v, u)}
\]

Where \(d(v,u)\) represents the shortest path distance between node \(v\) and \(u\),\(V\) is the set of all nodes in the graph.

We observed that, unlike Android malware\cite{intdroid,malscan}, Python malicious packages are often smaller in size\cite{sun2024}, and attackers tend to write malicious code directly in the global scope. As a result, even though the attackers use certain APIs, there are often no direct invocation relationships between them. This leads to a centrality value of zero for these APIs, This problem could result in a scenario where, although malicious packages invoke certain APIs, these invocation relationships are not accurately reflected in the centrality values. 
\begin{table}[htbp]
  \centering
  \caption{The feature set dimension after pre-processing by general-purpose LLM.}
\resizebox{\columnwidth
  }{!}{
    \begin{tabular}{|c|c|c|c|c|}
    \toprule
    Centrality & Closeness & Degree & Katz & Harmonic \\
    \midrule
    Total Dimensions  & 265 & 255 & 294 & 135 \\
    \bottomrule
    \end{tabular}%
    }
  \label{tab:set_dimension}%
\end{table}%
Consequently, important patterns of API usage may be overlooked. To address this issue, we adjust the calculation of centrality values by adding a default value of 1 to all centrality scores. This ensures no API's centrality value can be zero, effectively mitigating the problem and allowing us to capture the significance of APIs even in the absence of direct invocation relationships.

\subsection{Sensitive API Extraction and Filter}
After generating the API call graph and calculating node centrality values for each malicious package, the next step is to identify the sensitive API feature set. To achieve this, we aggregate the centrality values of APIs with the same name across all malicious packages. Specifically, we sum the centrality values of each API and then divide the total by the number of malicious packages to obtain an averaged centrality value for each API. This results in a comprehensive list of APIs with their corresponding averaged centrality values. We then rank the APIs based on their averaged centrality values and select the top \textit{K} (\(K=200, 300, 400, 500\)) as the sensitive API feature set. 
Although ML models can achieve effectiveness in malicious package detection comparable to that of LLMs, they have an inherent limitation: their inability to explain malicious behaviors. To address the need for a reliable dataset to explain malicious behaviors, we utilized an LLM (GPT-3.5-turbo) to construct a ground truth dataset. Specifically, the sensitive APIs extracted in Step 2 were analyzed using prompt engineering.

To ensure the feature set included as many suspicious APIs as possible while reducing the influence of irrelevant APIs, we selected the top 500 APIs for analysis. Each API was evaluated by the LLM to determine whether it could be used for malicious purposes: If potentially malicious, the API was retained in the feature set. Otherwise, it was removed. This process resulted in a refined, accurate, and relatively comprehensive feature set of suspicious APIs. The specific feature dimensions are summarized in Table~\ref{tab:set_dimension}.

In addition to fully leveraging the code understanding capabilities of the LLM, the filtered APIs were analyzed further. We instructed the model to generate outputs in a JSON format, similar to the gray-highlighted section in Figure~\ref{fig:overflow}, where each API was mapped to its potential malicious behaviors. After manual verification and the removal of unreasonable or inaccurate entries from the generated analysis, we finalized a ground-truth dataset that links API names to their potential malicious behaviors.

It is important to highlight that, unlike other approaches utilizing LLMs for malicious package detection, which require repeated invocations of the LLMs while our approach only necessitates a single invocation to construct the feature set. Additionally, the number of APIs that need to be analyzed is limited to 500. This results in significant advantages in terms of both time and economic costs.

\subsection{Malicious Package Detection}
To achieve efficient and precise detection of malicious packages, we utilized the feature set generated in Step 2 and applied methods such as regular expression matching to extract relevant features from the dataset's packages. For each API in the sensitive API feature set, we computed its corresponding centrality value within the context of a given package. This centrality value was then used as the feature value, reflecting the importance of the API within the package's structure. The extracted values were compiled into comprehensive feature vectors, which were subsequently used to train ML models.

\subsection{Explanation output based on LIME}
\begin{figure}
    \centering
    \includegraphics[width=\linewidth]{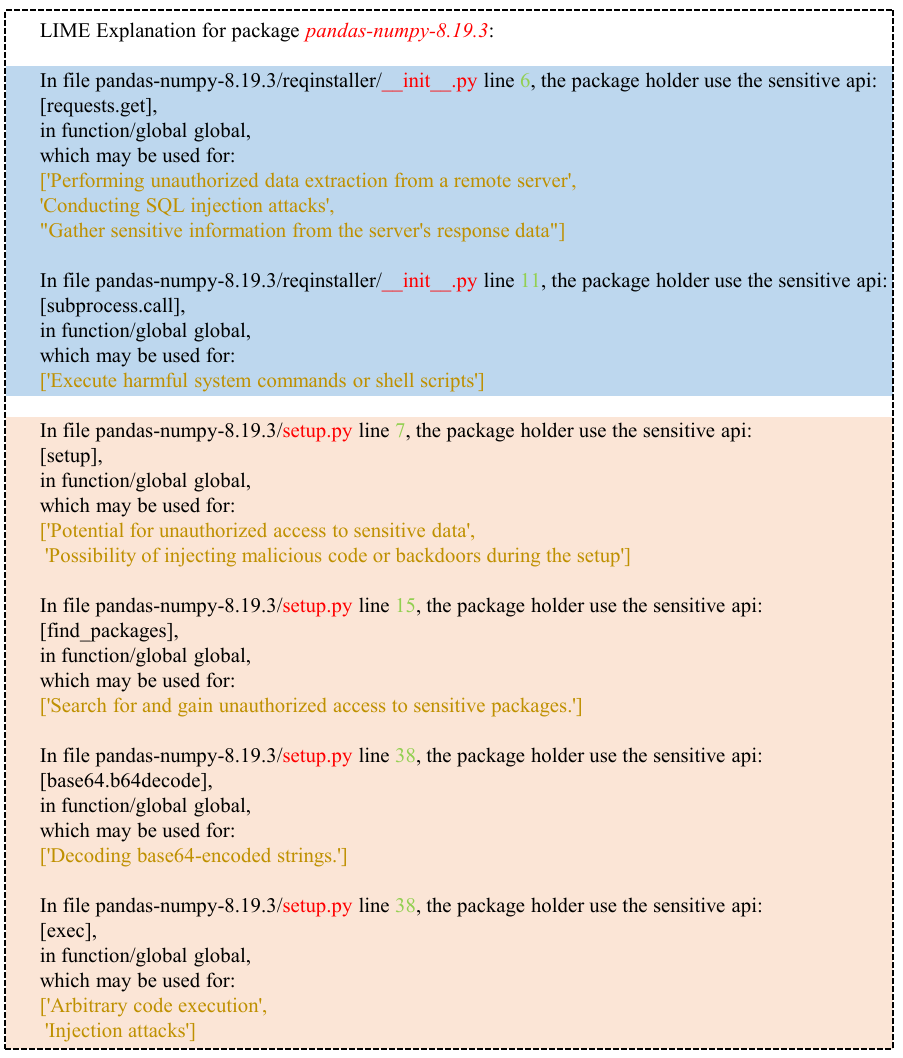}
    \caption{The explanation output result of malicious package \textit{pandas-numpy-8.19.3}.}
    \label{fig:explaintion}
\end{figure}

\noindent \textbf{Local Interpretable Model-agnostic Explanations (LIME)}. LIME is a widely used algorithm designed to enhance the explainability of ML models. It explains the predictions of any black-box model by approximating the model's behavior locally around a specific instance \cite{exp}. 
We integrated the LIME algorithm into our ML model to enhance its explainability by identifying the top 10 non-zero features that had the most significant impact on the model's predictions. For each of these features, we cross-referenced the API names with the ground truth dataset generated in Step 2 to retrieve all potential malicious behaviors associated with each API. Simultaneously, we located the specific lines of code where these APIs appeared within the analyzed package, allowing us to understand their exact usage. To provide a clearer view of the API interactions, we sorted the identified APIs based on the order in which they appeared in the code, creating a sequential representation of their calling order and relationships across different Python script files in the package. Finally, we output all suspicious APIs along with their respective potential malicious behaviors, providing a comprehensive and interpretable result for further investigation. The output results are shown in Figure~\ref{fig:explaintion}.

Each explanation output includes the following details: \textbf{Sensitive API Name:} The name of the sensitive API. \textbf{File Name and Code Line:} The name of the file in which the sensitive API is located, along with the specific line(s) of code where it is used. \textbf{Usage Context:} Whether the sensitive API is used in the global scope or within a specific function.
\textbf{Potential Malicious Behaviors:} A list of all possible malicious behaviors associated with the sensitive API.

\section{Evaluation}~\label{sec:eval}

In this section, we evaluate \system{} from different perspectives. First, we compare the effectiveness of \system{} with SOTA detection approaches (Section ~\cref{sec:eva_performance}). Second, we evaluate whether the feature set filtered by general-purpose LLMs can really affect the effectiveness of our approach (Section ~\cref{sec:eval_abalation}). Third, we evaluate if the explanation outputs generated by \system{} can truly help researchers capture the malicious intent of attackers (Section ~\cref{sec:eva_explanation}). Then we measure what is the optimal value for the parameter top $K$ in selecting sensitive APIs to achieve the best effectiveness of the model (Section ~\cref{sec:eva_parameter}). 
Besides, we evaluate the robustness of \system{} against the adversarial attack (Section~\cref{sec:adversarial}).
Finally, we show that \system{} can identify malicious packages that exist in the wild (Section ~\cref{sec:eva_paacticality}).

\PHB{Datasets and Models.} To better validate the effectiveness of \system{}, we constructed a model training and testing dataset. As shown in Table~\ref{tab:malicious}, we collected a dataset containing 9,664 malicious packages and 10,000 benign packages. The benign packages dataset was the same dataset mentioned in Section~\cref{sec:background_dataset}. The entire dataset was randomly divided into training and testing sets in an 8:2 ratio, with the former used for model training and the latter for testing.

\PHM{Baselines.} To evaluate the effectiveness of \system{} against existing approaches, we selected six SOTA approaches as our baselines for comparison. These six approaches are \textsc{VirusTotal}~\cite{virustotal}, \textsc{OSSGadget}~\cite{Microsoft2}, \textsc{Bandit4Mal}~\cite{FORK}, \textsc{Ea4mp}~\cite{sun2024}, \textsc{Cerebro}~\cite{DBLP:journals/corr/abs-2309-02637} and \textsc{GuardDog}~\cite{guarddog}. 
\textsc{VirusTotal}~\cite{virustotal} provides an online detection platform where packages can be uploaded directly for analysis. It automatically detects whether the software package contains suspicious files, IPs, URLs, \textit{etc.} 
\textsc{OSSGadget}~\cite{Microsoft2} can identify potential backdoors and malicious code within a package. 
\textsc{Bandit4Mal}~\cite{FORK} is an approach to finding common security issues in Python code. It processes each file, builds an AST from it, and runs appropriate plugins against the AST nodes. 
\textsc{Ea4mp}~\cite{sun2024} is an integrated detection approach based on deep code behavior sequences and metadata. Their approach used static analysis tools to extract CFGs and CGs to generate code behavior sequences, which were subsequently fine-tuned with the BERT model.  
\textsc{Cerebro} is an approach by extracts a code sequence that can describe the malicious behaviors of attackers, and then uses this sequence to fine-tune the BERT model.
\textsc{GuardDog} leverages Semgrep’s semantic analysis capabilities and YARA’s pattern-matching features to perform static code analysis and metadata scanning on packages from PyPI, NPM, and Go.

\PHB{Implementation.} We implement \system{} in Python using PyTorch \cite{PyTorch}. Our experiments are performed on a Linux workstation with an AMD RYZEN 7735HS CPU, 32GB RAM, and an NVIDIA V100 GPU with 32GB memory, running Ubuntu 22.04 with CUDA 12.1. We implemented the ML models using the scikit-learn library (sklearn) in Python. Specifically, we utilized five widely used classifiers: NB, XGBoost, RF, SVM, and MLP.

\PHB{Evaluation Metrics.} We utilize three widely-used binary classification metrics for evaluation: \textit{Precision}, \textit{Recall}, and \textit{F1 score}. 
\emph{Precision} is the ratio of correctly identified malicious samples to all samples classified as malicious, representing the model's ability to avoid false positives.  
\emph{Recall} measures the proportion of correctly identified malicious samples out of all actual malicious samples, reflecting the model's ability to detect true positives.  
\textit{F1 score} is the harmonic mean of \textit{Precision} and \textit{Recall}, providing a balanced measure of a model's effectiveness. It is calculated as follows: $2 \times \frac{Recall \times Precision}{Recall + Precision}$.

\begin{figure}
    \centering
    \includegraphics[width=\linewidth]{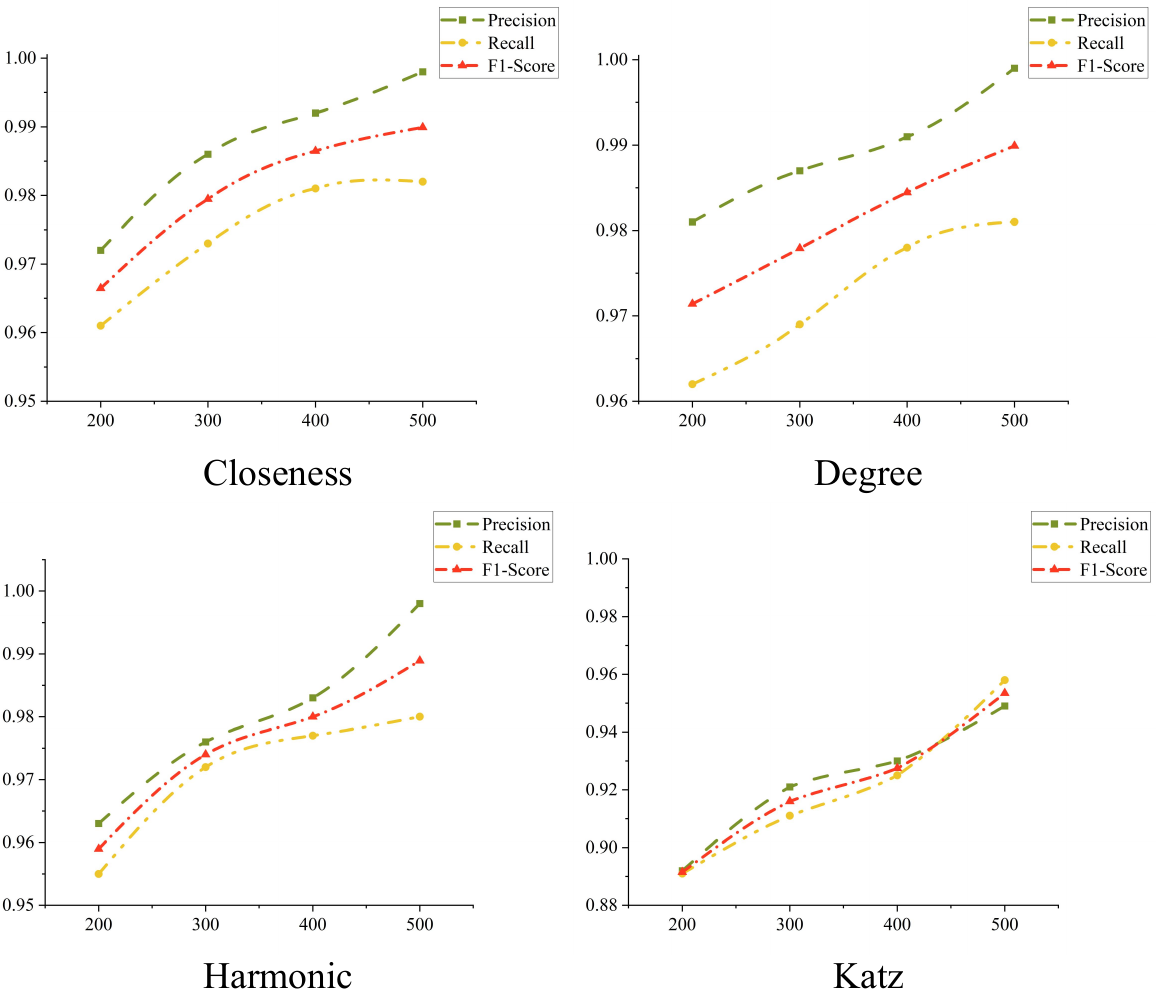}
    \caption{The effectiveness of the Random Forest (RF) model trained by top n APIs in four different centrality feature sets. }
    \label{fig:topn}
\end{figure}

\subsection{Effectiveness Evaluation}
\label{sec:eva_performance}
\PHM{Experimental Setup.} To evaluate the effectiveness of \system{}, we combined benign and malicious packages, then randomly split the dataset into two parts: 80\% was used as the training set, while the remaining 20\% served as the test set. The same dataset was used to evaluate \textsc{OSSGadget}, \textsc{Band4Mal}, \textsc{Ea4mp},\textsc{Cerebro} and ourapproach \system{}. Since \textsc{VirusTotal} provides an online detection platform. 
Similarly, \textsc{GuardDog} relies on static analysis and heuristic rules without requiring model training, so we only use the test set to verify their effectiveness.
The experimental results are shown in Table~\ref{tab:baseline_result}.

\PHM{Result.} Table~\ref{tab:baseline_result} shows the effectiveness of \system{} compared to the five baseline approaches on the same dataset. Comparing the data in the table, it is evident that \system{} improves precision by 0.5\% to 33.2\% over the other approaches, achieving the highest precision among all. Regarding recall, \system{} improved by 1.8\% to 22.1\% compared to all other approaches.
The experimental results demonstrate that our approach achieves optimal effectiveness in terms of \textit{precision}, \textit{recall}, and \textit{F1 scores}. In contrast, other approaches exhibit inherent limitations that affect their effectiveness, making them less effective compared to \system{}. \textbf{First,} Approaches like Bandit4Mal and OSSGadget often misclassify benign behaviors such as network connections and file operations in normal packages as malicious. This misclassification arises from their inability to distinguish between malicious and benign behaviors effectively. \textbf{Second,} Signature-based detection methods, such as VirusTotal, struggle to keep pace with the diversity and constant evolution of malicious code. The difficulty in maintaining consistent fingerprints for new malicious code makes it challenging for signature databases to detect the latest malicious packages. \textbf{Third,} Although LLMs often demonstrate excellent effectiveness in malicious package detection, as seen in approaches like \textsc{Ea4mp} and \textsc{Cerebro}, they tend to be overly sensitive to certain sensitive operations. This heightened sensitivity can lead to false positives, where benign packages are incorrectly flagged as malicious.

\noindent \PHM{Case Study.}
To understand why certain benign packages that also invoke sensitive APIs can be more effectively distinguished from malicious ones by our approach compared to existing methods, we analyzed benign samples that were misclassified by other approaches but correctly identified by ours. Since most existing methods function as black-box models, it is difficult to determine the rationale behind their decisions. In contrast, \textsc{GuardDog} provides a list of suspicious APIs contained in each flagged package, which offers partial insight into its classification process. Therefore, we selected benign samples misclassified by GuardDog for further analysis.

Our analysis revealed that \textsc{GuardDog} tends to exhibit strong sensitivity toward specific categories of APIs. Once such APIs are invoked in a package, \textsc{GuardDog} is more likely to classify it as malicious. For example, in the case of \textit{GACF-1.0.1}, the package invoked APIs such as \textit{os.environ.copy()}, \textit{subprocess}, and \textit{os.makedirs()}, which fall under the category of system-level operations that interact with the environment or the file system. These triggered a false positive detection by \textsc{GuardDog}. In contrast, our approach not only considers the presence of sensitive API calls but also incorporates their centrality values within the API call graph as feature indicators. We found that, except for the relatively high centrality of the subprocess API, the other sensitive APIs in this benign package exhibited no significant centrality anomalies. This suggests that the benign package did not rely heavily on a small set of sensitive APIs, as reflected in their low centrality values. This result demonstrates that incorporating API centrality can help reduce false positives and more accurately distinguish benign packages from malicious ones, outperforming existing approaches.

\begin{table}[htbp]
  \centering
  \caption{Effectiveness comparison with the SOTA baselines.}
  \resizebox{\columnwidth}{!}{
    \begin{tabular}{c|ccc}
    \toprule
    Approach & Precision (\%) & Recall (\%) & F1 score (\%) \\
    \midrule
    \textsc{VirusTotal}~\cite{virustotal} & 95.2  & 80.6  & 87.3  \\
    \textsc{OSSGadget}~\cite{Microsoft2} & 74.8  & 85.0  & 79.6  \\
    \textsc{Band4Mal}~\cite{FORK} & 84.8  & 96.7  & 90.4  \\
    \textsc{Ea4mp}~\cite{sun2024} & 99.1  & 95.4  & 97.2  \\
    \textsc{Cerebro}~\cite{DBLP:journals/corr/abs-2309-02637} & 98.6  & 85.7  & 91.7 \\
    \textsc{GuardDog}~\cite{guarddog} & 95.6  & 82.6  & 88.6 \\
    \midrule
    \system{} & \textbf{99.6}  & \textbf{98.4}  & \textbf{99.0}  \\
    \bottomrule
    \end{tabular}%
    }
  \label{tab:baseline_result}%
\end{table}%

\begin{table*}[htbp]
  \centering
  \caption{Effectiveness of models trained by different centrality feature sets.}
  \resizebox{2\columnwidth
  }{!}{
    \begin{tabular}{|c|c|cccc|cccc|}
    \toprule
    \multicolumn{1}{|r}{} & \multicolumn{1}{r}{} & \multicolumn{4}{c|}{with Feature Filtering} & \multicolumn{4}{c|}{w/o Feature Filtering} \\
    \midrule
    \multicolumn{2}{|c|}{Metrics (\%)} & Closeness & Harmonic & Degree & Katz & Closeness & Harmonic & Degree & Katz \\
    \midrule
    \multirow{3}[6]{*}{RF} & Precision & 99.4  & 92.5  & 99.3  & 99.6  & 99.9  & 99.9  & 99.9  & 94.9  \\
\cmidrule{2-2}      & Recall & 97.0  & 97.1  & 97.3  & 98.4  & 98.1  & 98.0  & 98.2  & 95.8  \\
\cmidrule{2-2}      & F-1 & 98.2  & 94.8  & 98.3  & 99.0  & 99.0  & 99.0  & 99.1  & 95.3  \\
    \midrule
    \multirow{3}[6]{*}{XGBoost} & Precision & 99.4  & 99.2  & 92.5  & 99.3  & 99.2  & 99.3  & 99.2  & 93.0  \\
\cmidrule{2-2}      & Recall & 96.5  & 96.3  & 95.5  & 96.9  & 98.5  & 98.7  & 98.6  & 94.5  \\
\cmidrule{2-2}      & F-1 & 97.9  & 97.7  & 94.0  & 98.1  & 98.8  & 99.0  & 98.9  & 93.7  \\
    \midrule
    \multirow{3}[6]{*}{SVM} & Precision & 97.9  & 87.1  & 97.6  & 97.9  & 82.8  & 86.6  & 72.6  & 71.8  \\
\cmidrule{2-2}      & Recall & 96.5  & 91.2  & 96.2  & 96.3  & 80.9  & 83.5  & 96.1  & 95.9  \\
\cmidrule{2-2}      & F-1 & 97.2  & 89.1  & 96.9  & 97.1  & 81.8  & 85.0  & 82.7  & 82.1  \\
    \midrule
    \multirow{3}[6]{*}{MLP} & Precision & 98.5  & 92.0  & 98.2  & 98.4  & 99.0  & 99.1  & 98.3  & 89.8  \\
\cmidrule{2-2}      & Recall & 97.8  & 97.0  & 98.0  & 98.1  & 98.9  & 95.5  & 98.6  & 92.5  \\
\cmidrule{2-2}      & F-1 & 98.1  & 94.4  & 98.1  & 98.2  & 99.0  & 97.3  & 98.4  & 91.1  \\
    \bottomrule
    \end{tabular}%
    }
  \label{tab:result2}%
\end{table*}%

\subsection{Ablation Study}
\label{sec:eval_abalation}

\noindent \PHM{Experimental Setup.} To validate the effectiveness of using LLMs for feature selection and analysis, we designed a comparative experiment to assess the impact of two different feature sets on the experimental results. For the initial feature set, we selected the top 500 APIs ranked by centrality values as sensitive APIs and used this feature set for feature extraction from the dataset. For the second feature set, we refined the initial set by utilizing the \textit{GPT-3.5-turbo} model for further filtering and analysis. To comprehensively evaluate the classification effectiveness of the model on both benign and malicious packages, we separately calculated the \textit{Precision}, \textit{Recall}, and \textit{F1 score} for benign and malicious samples.

\PHM{Result.} Through the analysis of APIs filtered out by the large language model, we observed that APIs like \textit{print}, \textit{range}, and \textit{int}, despite being ranked within the top 500, are primarily used for basic data processing or output operations. These APIs are seldom, if at all, employed as vehicles for malicious activities.
The experimental results, shown in Table~\ref{tab:result2}, reveal that for the NB model, simply selecting the top 500 APIs as the feature set causes the model to overly favor classifying packages as either entirely benign or entirely malicious, rendering it almost ineffective in practical applications. However, the feature set refined using the LLM significantly improved the NB model's effectiveness. While its effectiveness still lags behind that of other models, this is primarily due to the inherent limitations of the NB model itself.  
The experimental results show that the remaining four ML models performed similarly across both feature sets. These findings support two key conclusions:
\textbf{First}, our approach of using API call graph centrality for automated feature extraction is effective. This indicates that centrality measures can successfully capture relevant features for malicious package detection.
\textbf{Second}, leveraging a general large language model can effectively help in filtering out irrelevant APIs from the feature set.

\subsection{Explainability Evaluation}
\label{sec:eva_explanation}

\PHM{Explanation outputs verification dataset.}
To validate the accuracy of the explanation outputs, we randomly selected 100 malicious packages and 100 benign packages from the dataset for analysis. First, we employed prompt engineering to query the \textit{GPT-3.5-turbo} model, instructing it to generate malicious behavior analyses in a specified format for the selected packages. To mitigate the potential impact of hallucinations in the LLM on the experimental results, we further conducted manual verification of the model's outputs. This process resulted in the creation of an explanation output verification dataset that contains 100 malicious packages.

To validate the accuracy of the model's final explanation outputs, we used the \textbf{Explanation Outputs Verification Dataset} as a benchmark. Two evaluation criteria were adopted: the number of sensitive APIs included in the output and their precise localization. If an explanation output contained at least 80\% of the sensitive APIs and correctly identified their locations, it was deemed accurate.   

The experimental results, shown in Table~\ref{tab:expalanation_result}, all four ML models successfully identified the vast majority of malicious packages and generated accurate explanation outputs. Specifically, for degree centrality, 96 malicious packages were detected by at least three models, and 90 were identified by all four models. Even for the feature set based on harmonic centrality, which performed less effectively, there were 79 malicious packages detected by the four models.

\begin{figure}
    \centering
    \includegraphics[width=\linewidth]{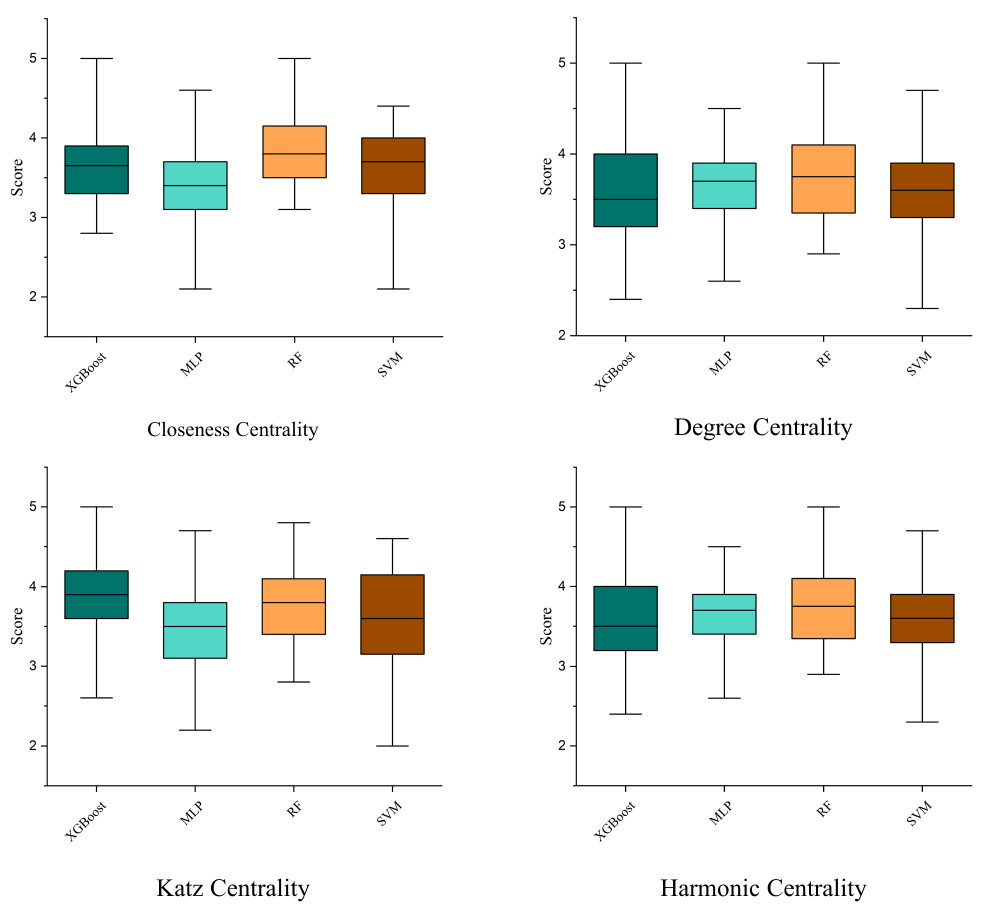}
    \caption{Box plot analysis of the average score of 100 malicious packages’ explanation output.}
    \label{fig:score}
\end{figure}

To further assess whether the explanation output could help researchers analyze malicious behaviors, we randomly invited \textit{n=24} volunteers (each with at least two years of experience in software engineering or software security) to rate the explanation results. Ratings ranged from 1 to 5, with higher scores indicating better quality. If a model failed to detect a malicious package or did not generate any explanation output, a score of 0 was assigned. The final scores were averaged and shown in Figure~\ref{fig:score}.  

\begin{table}[htbp]
  \centering
  \caption{Effectiveness of different ML Models in \textbf{Explanation Outputs Verification Dataset} (The Third Column shows the number of malicious packages that every model can detect and explain while the Fourth and Fifth Columns show the number of malicious packages that can be detected and accurately explained by more than 3 or 4 different models.).}
  \resizebox{0.8\columnwidth
  }{!}{
    \begin{tabular}{|c|c|c|c|c|}
    \toprule
    Centrality  & Model & Total detected & r>=3 & r=4 \\
    \midrule
    \multirow{4}[10]{*}{Closeness} & XGBoost & 96 & \multirow{4}[10]{*}{95} & \multirow{4}[10]{*}{93} \\
\cmidrule{2-3}      & RF & 95 &   &  \\
\cmidrule{2-3}      & SVM & 94 &   &  \\
\cmidrule{2-3}      & MLP & 98 &   &  \\
    \midrule
    \multirow{4}[10]{*}{Degree} & XGBoost & 97 & \multirow{4}[10]{*}{96} & \multirow{4}[10]{*}{90} \\
\cmidrule{2-3}      & RF & 96 &   &  \\
\cmidrule{2-3}      & SVM & 91 &   &  \\
\cmidrule{2-3}      & MLP & 97 &   &  \\
    \midrule
    \multirow{4}[10]{*}{Katz} & XGBoost & 95 & \multirow{4}[10]{*}{93} & \multirow{4}[10]{*}{86} \\
\cmidrule{2-3}      & RF & 93 &   &  \\
\cmidrule{2-3}      & SVM & 88 &   &  \\
\cmidrule{2-3}      & MLP & 96 &   &  \\
    \midrule
    \multirow{4}[10]{*}{Harmonic} & XGBoost & 95 & \multirow{4}[10]{*}{92} & \multirow{4}[10]{*}{79} \\
\cmidrule{2-3}      & RF & 93 &   &  \\
\cmidrule{2-3}      & SVM & 82 &   &  \\
\cmidrule{2-3}      & MLP & 95 &   &  \\
    \bottomrule
    \end{tabular}%
    }
  \label{tab:expalanation_result}%
\end{table}%

The experimental results demonstrate that the explainability content generated by our approach achieved an average score of 3.5 or higher, indicating that the explanation outputs are effective and useful for aiding in malicious behavior analysis.

\noindent \textbf{False Positive Analysis.}
To investigate why the model misclassified certain benign samples as malicious, we conducted a detailed analysis of three false positive cases identified by the Random Forest model (e.g., \textit{eazure-0.1.1}, \textit{gitlab-clone-0.1.1}, \textit{py-sourcemap-.1.14}) and examined the corresponding explainability outputs. 
We found that these benign packages all invoked multiple high-ranking sensitive APIs from the feature set. 
For example, \textit{py-sourcemap-0.1.14} performed file operations (e.g., \textit{read}, \textit{open}), issued network requests (e.g., \textit{urlopen}), and executed installation-related commands (e.g., \textit{install.run}). 
These APIs also showed significantly higher centrality values within the package’s API call graph. 
These findings suggest that the frequent use of diverse high centrality sensitive APIs likely led to the model’s misclassification.

\subsection{Hyperparameter Sensitivity Analysis}
\label{sec:eva_parameter}
\noindent \PHM{Experiment Setup.} To systematically examine the impact of varying the top $K$ parameter on the final model's effectiveness, we performed experiments with different ($K$ \(= 200, 300, 400, 500 \)). The lower bound of $K$ \( = 200 \) was established based on the dimensionality of a manually curated feature set, which consisted of 132 features, ensuring that the automatically extracted feature set would not be less informative. The upper bound of $K$\( = 500 \) was derived from manual analysis, which demonstrated that APIs ranked beyond this threshold seldom exhibited suspicious or malicious behavior. We selected the Random Forest (RF) model that demonstrated the best effectiveness in our previous experiments to evaluate the effect of these parameter choices on model effectiveness. This approach facilitates a thorough evaluation of how the selection of $K$ influences both the effectiveness and reliability of the detection system, thereby providing insights into optimal parameter configuration for malicious package detection.

\noindent \PHM{Result.} The experimental results, illustrated in Figure 2, show that as $K$ increases, the model's effectiveness consistently improves across feature sets derived using four different centrality metrics. For instance, the \textit{F1 scores} increase by 2\%–7\% when $K$ is raised from 200 to 500, indicating that higher $K$ values include more suspicious APIs in the feature set. These findings suggest that setting $K$\( = 500 \) allows the feature set to capture the most comprehensive set of suspicious APIs. To ensure optimal model effectiveness, all subsequent experiments adopt $K$\( = 500 \) as the default parameter setting.

\subsection{Robustness against Adversarial Attack}
\label{sec:adversarial}
\PHM{Experiment Set.} To evaluate the robustness of \system{} against adversarial attacks, we selected two representative and widely adopted categories of attack strategies~\cite{zhangfighting}.
\textbf{Category 1: Feature Space Attacks.} This category targets the feature vectors that are used by our detection model. Two distinct attack methods were applied: The first method introduces random noise into the feature vectors, following the randomization-based attack~\cite{xie2017mitigating}. The second method transforms all non-zero feature values in both the training and test sets to 1, instead of using their original centrality scores of API nodes.
\textbf{Category 2: Source Code Level Adversarial Attack.} 
Several adversarial attack strategies against software packages have been proposed in prior works. For example, Kreuk~\cite{DBLP:journals/corr/abs-1802-04528} injects adversarial byte sequences into binary files; IPR~\cite{10.1145/2897845.2897863} obfuscates code by inserting randomized, semantically ineffective instructions, and DISP~\cite{DBLP:conf/sp/PappasPK12} introduces code randomization using equivalent instruction replacement.
However, these techniques target executable binaries (e.g., \textit{.exe}, \textit{.apk}), whereas most malicious PyPI packages are distributed as source archives (e.g., \textit{.tar.gz}). Moreover, to ensure the effectiveness of static analysis tools used to construct API call graphs, the injected content must preserve source code validity. Therefore, we designed a \textbf{source code level attack} inspired by IPR~\cite{10.1145/2897845.2897863}, adapted to  Python packages. 
\relatedrv{Specifically, for each malicious package, we randomly selected $\alpha$ benign packages (\textit{1 <= $\alpha$ <= 3}). From these, we randomly chose $\beta$ Python source code files to inject. After analyzing the dataset of malicious packages, we found that each package contains an average of 4.18 \textit{.py} files. To avoid injecting an excessive amount of dead code that could distort the package structure, we rounded the value $\beta$ up to a maximum of 5. To maintain randomness in the poisoning process, we allowed $\beta$ to vary within the range of 1 to 5.}
These benign files were then injected into the malicious package without modifying any original code. Then we reconstructed the API call graph, re-extracted the feature set and feature vectors based on closeness centrality, and retrained the detection models.

\begin{figure}
    \centering
    \includegraphics[width=\linewidth]{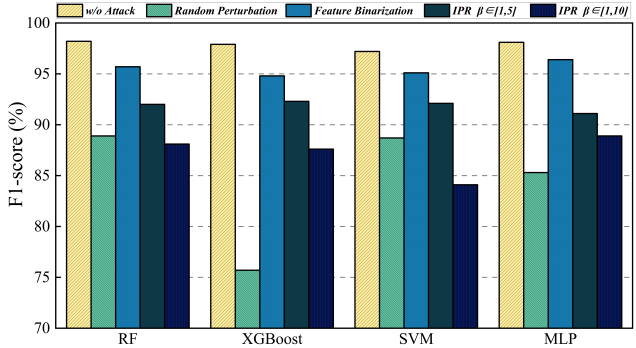}
    \caption{\relatedrv{Model robustness against adversarial attacks.}}
    \label{fig:attack}
\end{figure}

\noindent \PHM{Result.} Figure~\ref{fig:attack} presents the \textit{F1 score} of the four ML models used in our approach (RF, XGBoost, SVM, and MLP) when subjected to adversarial attacks.
\textbf{Under the first category of attacks,} we observed that the strategy of replacing all non-zero feature values with \textit{1} led to only a slight drop in effectiveness. This is expected, as the transformation does not alter the underlying API invocation patterns of the malicious packages. In contrast, the randomization-based strategy introduces noise to the feature vectors, effectively disrupting the API call patterns. As a result, the \textit{F1 scores} dropped more substantially. Nevertheless, even under this more aggressive attack, the worst-performing model (XGBoost) still achieved an \textit{F1 score} of 75.7\%, while the remaining models maintained \textit{F1 scores} above 85\%.
\textbf{Under the second category of attacks}, our approach also experienced a effectiveness decline. However, even for the worst-performing model (MLP), the \textit{F1 scores} remained as high as 91.1\%. 
\relatedrv{Even when we increased the parameter $\beta$ from 1 to 10, our approach was still able to maintain an \textit{F1 score} above 84\%.}
Additionally, the effectiveness gap among the four models under this IPR-inspired source-level attack was relatively small.

To understand this result, we analyzed the post-attack feature sets and found that although the injected benign code (i.e., dead code) somewhat reduced the centrality values of certain sensitive APIs, these added components did not form direct invocation relationships with the sensitive APIs. As a result, within the local subgraphs containing sensitive APIs, their centrality remained significantly higher than that of unrelated APIs. These findings demonstrate that although our approach suffers some degradation under adversarial conditions, it maintains effectiveness at an acceptable level, highlighting its robustness against such attacks.

\subsection{Practicality}
\label{sec:eva_paacticality}
\PHM{Wild truth.}
To verify if \system{} can identify malicious packages that exist in the wild, we crawled \emph{all} packages uploaded to PyPI between December 21, 2024, and January 28, 2025, from the official website~\cite{PYPISample}. In total, we collected 64,348 packages for real-world validation. Two authors separately review the reported malicious packages. All suspicious packages (including samples that did not reach a consensus) would be forwarded to a security expert from a prominent IT enterprise with at least five years of experience in software supply security to conduct a secondary review.

In total, \system{} discovered 144 suspicious packages. After manual review, 113 out of them were confirmed malicious. We reported these packages to the PyPI official. As of January 21, 2025, 109 of them have been removed.

\noindent \PHM{False Positives.} Upon analyzing 31 packages that were incorrectly flagged as malicious, we discovered that 25 of them were user-uploaded test demos. Although these demo packages utilized numerous suspicious APIs, such as \textit{request.get} and \textit{subprocess.run}, they did not engage in any malicious behavior. The remaining 6 packages were identified as prank packages, such as the \textit{"crazy-thursday"} package, which creates local processes to display "Crazy Thursday" messages to users but does not perform any harmful actions. We have reported these prank packages to the PyPI official too.

\section{Discussion}~\label{sec:discuss}

\noindent \PHM{Code Obfuscation.} Code obfuscation is a common technique used to evade existing detection approaches. Currently, most approaches primarily analyze software packages based on their source code files. Some attackers circumvent detection by packaging their code into binary executable files. Detecting such packages requires software security professionals to have reverse engineering skills and to continuously monitor the resource usage of the software package during execution.
Fortunately, code obfuscation also requires some technical expertise from the attackers. Currently, the majority of malicious software packages mainly obfuscate the parameter values of functions. This means that \system{} can still accurately detect them. However, how we deobfuscate more complex forms of code obfuscation remains a challenge that we need to address.

\noindent \PHM{Cross Platforms Detection.} Although existing approaches for detecting malicious packages demonstrate strong effectiveness, most of these methods are tailored to one or two specific platforms. In the broader context of the open-source software ecosystem, each programming language has its own maintained open-source community. This implies that for different open-source communities, maintainers need to employ distinct analysis methods and train separate detection models. Such a requirement undoubtedly increases the complexity of maintaining the stability of these communities. Zhang et al.~\cite{DBLP:journals/corr/abs-2309-02637} proposed a dual-platform (NPM\&PyPI) detection tool based on the BERT model and code behavior sequences. Their experiments demonstrated that although API names vary across platforms, malicious actors must invoke APIs with specific functionalities. Given this, the intrinsic relationships among these APIs may potentially be leveraged to develop cross-platform detection approaches for malicious packages.

\section{Related Work}
\label{sec:ralated}

\noindent \PHM{Malicious Package Detection.} 
Detecting malicious software packages within open-source software registries is a complex challenge. Liang et al.~\cite{DBLP:conf/trustcom/LiangZWDH21} introduced PPD, a third-party malware identification framework employing anomaly detection. This approach forms a comprehensive code package by importing required packages, uses AST (Abstract Syntax Tree) and RegExp (Regular Expressions) to extract code features (e.g., IP addresses, dangerous functions), and incorporates the Levenshtein distance of package names into the feature set. It then applies anomaly detection algorithms to identify malicious packages.  Given that developers often host open-source code on platforms like GitHub, inconsistencies between the code released on registries such as PyPI and the corresponding GitHub repositories may signal malicious injection. To address this issue, Vu et al.~\cite{DBLP:conf/sigsoft/VuMPPS21} proposed \textsc{LastPyMile}, a framework designed to identify disparities between software package construction artifacts and their source repositories. This approach enhances monitoring of registry security, helping mitigate risks. Zhang et al.~\cite{DBLP:journals/corr/abs-2309-02637} proposed \textsc{Cerebro}, which extracts code behavior sequences based on abstract syntax trees. By identifying available APIs in the AST, they construct a sequence describing the attacker's malicious behavior and use this sequence to fine-tune a BERT model. However, while Zhang's method leverages code behavior sequences to enhance model understanding of attack patterns, it remains limited by the reliance on manual feature recognition. Liang et al.~\cite{DBLP:conf/kbse/LiangLWLW23} introduced \textsc{MPHunter}, which identifies malicious packages by extracting code behavior sequences, converting them into vectors, and employing clustering to detect anomalies. However, \textsc{MPHunter} is constrained to analyzing only the `setup.py` script in packages. As noted by Guo et al.~\cite{DBLP:journals/corr/abs-2309-11021}, attackers often distribute malicious code across multiple scripts, complicating detection and circumventing such approaches.

\noindent \PHM{Package Registry Security.} There are numerous repositories exploited as platforms for distributing malicious code and software libraries. 
\textsc{GuradDog}~\cite{guarddog} is an open-source command-line tool (CLI) developed by Datadog, designed to identify malicious packages in PyPI, npm, and Go through static code analysis and metadata scanning. It combines the semantic analysis capabilities of Semgrep with the pattern-matching power of YARA to detect suspicious behaviors.
\textsc{Anomalicious}~\cite{DBLP:conf/icse/Gonzalez0GS21} addresses this issue by leveraging commit logs and repository metadata to automatically identify anomalies and potentially malicious commits. Attackers often misuse GitHub's fork functionality to store and distribute malware~\cite{DBLP:conf/icse/Gonzalez0GS21}.
To counter this threat, Zhang et al.~\cite{DBLP:conf/icbk/ZhangFHYXLSZX20} proposed an enhanced deep neural network (DNN) \cite{BGNN4VD,MVD} for analyzing the code content of GitHub repositories. Their approach employed a heterogeneous information network (HIN) to model neighborhood relationships, thereby improving recognition accuracy. Malicious actors frequently embed harmful shell commands within Python scripts to achieve illicit objectives. 
Traditional static analysis methods often struggle to detect such sophisticated attacks. To address this gap, Zhou et al.~\cite{DBLP:journals/jifs/ZhouHHLS22} introduced \textsc{PyComm}, ML model specifically designed to detect malicious commands in Python scripts. \textsc{PyComm} evaluates multidimensional features, simultaneously analyzing 12 statistical characteristics of Python source code and string sequences. In addition, Fang et al.~\cite{DBLP:journals/scn/FangXH21} employed ML techniques to identify Python backdoors. Their method represented text using statistical features derived from obfuscation and opcode sequence characteristics during compilation. By matching suspicious modules and functions within the code, their approach effectively detected embedded backdoors.
\section{Conclusion}~\label{sec:conclusion}

In this paper, we demonstrate that with a sufficiently comprehensive feature set, ML models can achieve effectiveness comparable to that of LLMs. We also proposed a novel approach \system{}, leveraging graph centrality to extract the sensitive APIs automatically, eliminating reliance on manually predefined feature sets. Moreover, we employ GPT-3.5-turbo to refine and analyze the feature set, and by integrating the LIME algorithm, we achieved explainable outputs for malicious package detection, enhancing both explainability and effectiveness. We evaluate \system{} against the state-of-the-art approaches on a newly-constructed dataset. \system{} performs better on all metrics. 
Moreover, compared to existing works, our approach supports much faster iteration. The computation of centrality values for all malicious packages can be completed within a few hours, while feature extraction and model training require only a few minutes. This efficiency enables our approach can be updated on a daily basis using newly emerging packages, ensuring timely adaptation to evolving threats.
We also applied \system{} to real-world detection and found 113 malicious packages by detecting 64,348 newly uploaded PyPI packages, 109 of which have been removed by PyPI officials. This indicates that \system{} is a practical approach that can be adopted by the Python community to detect malicious packages.

\section*{Acknowledgments}

We thank the reviewers for their valuable comments. This work was supported in part by the National Natural Science Foundation of China (Grant No. 62402342), the Jiangsu “333” Project, Postgraduate Research \& Practice Innovation Program of Jiangsu Province (KYCX24\_3747), and Shanghai Sailing Program (No. 24YF2749500).

\clearpage

\section*{Ethics Considerations}~\label{sec:ethics}


Throughout the study, we ensured that no personally identifiable information (PII), sensitive user data was involved at any stage of data collection, analysis, or validation. All datasets used for training and testing, including benign and malicious packages, were obtained from publicly available repositories or previous peer-reviewed studies (\textit{e.g.,} ~\cite{DBLP:conf/trustcom/LiangZWDH21, DBLP:journals/corr/abs-2309-02637, DBLP:conf/kbse/LiangLWLW23, DBLP:journals/corr/abs-2309-11021, DBLP:conf/icse/Gonzalez0GS21, DBLP:journals/scn/FangXH21}).

To ensure ethical integrity, we followed a transparent and responsible disclosure procedure for all identified malicious packages. Specifically, when our tool, \textsc{MalGuard}, detected previously unknown malicious packages among newly uploaded PyPI packages, we reported these findings to the PyPI security team prior to any public disclosure. Out of 113 packages flagged as malicious, 109 were subsequently reviewed and removed by PyPI, indicating the practical value and responsible execution of our methodology.

\section*{Open Science}
This work aligns with the principles of Open Science and aims to facilitate transparency, reproducibility, and community collaboration.
All resources are available via our project repository at: \url{https://doi.org/10.5281/zenodo.15545824}. We provide full documentation and guidelines to replicate our experiments.

\bibliographystyle{plain}
\bibliography{main}

\appendix
\section*{Appendix}
\section{LLM Prompts for Malicious Analysis}
Here are three prompts that we used to query LLM for malicious analysis. The first one is used to analyze whether the package that we send is malicious, illustrated in \textit{System Role Prompt of Malicious Packages Detection.} The second one is used to analyze weather the API we send can be used for malicious purposes, if so, analyze the malicious purposes, illustrated in \textit{System Role Prompt of Sensitive API Analysis.} The second one is used to analyze which APIs the malicious package we send has used, and describe the malicious behavior it contains, illustrated in \textit{System Role Prompt of Malicious Packages Analysis.}

\begin{tcolorbox}[title=System Role Prompt of Malicious Packages Detection]
\myparatight{Task:}

You are a cybersecurity expert. Your task is to analyze a given Python code snippet and determine whether it contains malicious behavior.

\myparatight{Guidelines:}

-Examine the code for patterns commonly used in malware.

-Consider whether the code could be harmful when executed, even if it appears simple.

-Be objective. If malicious behavior is suspected, clearly explain why.

-Output must be in valid JSON format.

\myparatight{Code:}

<INSERT CODE HERE>

\myparatight{JSON Response:} 
\begin{verbatim}
{
  "is_malicious": true or false,
  "reason": “The malicious behaviors the 
             code cotained.",
  "confidence": 0-1,
  "indicators": ["", ""]
}
\end{verbatim}

\end{tcolorbox}

\begin{tcolorbox}[title=System Role Prompt of Sensitive API Analysis]
\myparatight{Task:}

You are a security API auditor. Your task is to determine whether a given Python API can potentially be used for malicious purposes.

\myparatight{Guidelines:}

-Consider common attack techniques such as command execution, code obfuscation, data exfiltration, privilege escalation, etc.

-If the API is not typically used in a malicious context, return a neutral evaluation.

-Output must follow the required JSON format.

\myparatight{Code:}

<INSERT CODE HERE>

\myparatight{JSON Response:} 
\begin{verbatim}
{
  "api_name": “The name of the API",
  "is_potentially_malicious": true,
  "malicious_usage": “Malicious purposes."
}
\end{verbatim}

\end{tcolorbox}

\begin{tcolorbox}[title=System Role Prompt of Malicious Packages Analysis]
\myparatight{Task:}

You are a behavioral malware analyst. Your task is to extract and classify all malicious behaviors present in a given Python code snippet.

\myparatight{Guidelines:}

-Identify and label all suspicious or clearly malicious behaviors in the code.

-For each behavior, provide a category, a short description, and a relevant code snippet.

-If no malicious behaviors are found, return an empty array.

-Output must be in valid JSON format.

\myparatight{Code:}

<INSERT CODE HERE>

\myparatight{JSON Response:} 
\begin{verbatim}
{
  "malicious_behaviors": [
  {
    "type": “Malicious type n",
    "description": “Malicious behavior 
                        description.",
    "code_snippet": “The code that contain 
                          malicious APIs."
  },
  ...]
}
\end{verbatim}

\end{tcolorbox}

\end{document}